\documentclass[twocolumn,showpacs,preprintnumbers,amsmath,amssymb]{revtex4}

\usepackage{enumitem}
\usepackage{latexsym}
\usepackage{graphics}
\usepackage{amssymb}
\usepackage{bbm}
\usepackage{dcolumn}
\usepackage[dvips]{graphicx}
\usepackage[dvips]{color}

\def\sp{{\sigma_+}}
\def\sm{{\sigma_-}}
\def\po{\psi_{1}}
\def\pt{\psi_{2}}
\def\go{\gamma_1}
\def\gt{\gamma_2}

\begin{document}

\title{Stochastic Wave-Function Unravelling of the Generalized Lindblad Master Equation}
\author{Mervlyn Moodley}
 \email{moodleym2@ukzn.ac.za}
\affiliation{School of Physics, Quantum Research Group, University of KwaZulu-Natal, Westville Campus, Private Bag X54001, Durban, 4000, South Africa}
\author{Francesco Petruccione}
 \email{petruccione@ukzn.ac.za}
\affiliation{School of Physics, Quantum Research Group, University of KwaZulu-Natal, Westville Campus, Private Bag X54001, Durban, 4000, South Africa}
\affiliation{National Institute for Theoretical Physics, University of KwaZulu-Natal, Westville Campus, Private Bag X54001, Durban, 4000, South Africa}
\date{\today}

\begin{abstract}
Recently a generalized master equation was derived that extends the Lindblad theory to highly non-Markovian quantum processes (H.-P. Breuer, Phys. Rev. A \textbf{75}, 022103 (2007)). We perform a stochastic unravelling of this master equation by considering $n$ random state vectors that satisfy the corresponding stochastic differential equation for a piecewise deterministic process. As an application we consider a two-state system randomly coupled to an environment consisting of two energy bands with finite number of levels. Our numerical results are compared to results obtained from the time-convolutionless (TCL) projection operator method using correlated projection superoperators and the exact solution of the Schr\"{o}dinger equation for this system.
\end{abstract}

\maketitle

\section{\label{sec:level1}Introduction}

The success of present and future quantum technologies relies almost entirely on the quantum device's interaction with the environment it is in. Decoherence and dissipation phenomena dictate how much information can be transmitted from one quantum manipulation to the next. Decoherence, which is the loss of phase coherence between superpositions of quantum states, and dissipation, which is the leakage of population from the system to the environment, are major hurdles to the realization of realistic quantum technologies. As a result, the investigation of the dynamics of open quantum systems, is of utmost importance to our understanding of such undesirable phenomena~\cite{BP}.

Most approaches to the investigation of open quantum systems are based on Markovian assumptions, which makes use of the Born and Markov approximations that ultimately lead to the quantum Markov equation in Lindblad form~\cite{Lind,GKS}. In most cases this Lindblad master equation is stochastically unravelled enabling the efficient use of stochastic wave function methods to analyze the dynamics of the open quantum system. These methods have prominence in applications to many quantum optical systems~\cite{DCM,GP,CM,CM2,MC}.

In some instances however, open quantum systems associated with more realistic quantum technological process are classified as non-Markovian. Some prime indicators of the presence of non-Markovian effects and the failure of Markovian approximations are when the system-environment couplings are strong or when the initial states are classically correlated or entangled. Some examples of non-Markovian systems include spin star systems~\cite{BP1,BBP} and circuit QED~\cite{CQED1, CQED2}. Various techniques have been developed to describe non-Markovian quantum process. Generalized or non-Markovian master equations have first been introduced in Refs.~\cite{EG,Bu}. The Nakajima-Zwanzig formalism~\cite{Na,Zw} and the time-convolutionless projection operator method~\cite{STH,CS,SA} have proved to be useful in deriving approximations based on projection operator techniques. The latter method, employing correlated projection superoperators, was recently used to derive a non-Markovian generalization of the Lindblad equation~\cite{B1}. Stochastic wave function methods have also been proposed and developed for non-Markovian quantum master equations~\cite{HKP,B2,GAW} and more recently by Piilo \textit{et al}~\cite{Pii}.

In this paper we perform a stochastic unravelling of the generalized Lindblad master equation which allows for the use of traditional Markovian stochastic wave-function simulations. This approach is applicable to both time dependent and time independent rates. As an application we consider a two-level system coupled to an environment consisting of two energy bands, each with a large number of energy levels. Due to its highly non-Markovian characteristic, this model has gained some interest over the past couple of years~\cite{GM,BGM,B1,HY}. In Ref.~\cite{BGM}, the time-convolutionless projector operator technique and the Hilbert-space-average method was used to analyze this model; our Monte Carlo simulations are compared to the former technique. Similar models of this type have also been studied before. These include the model by Esposito and Gaspard~\cite{EGG}, and the models by Bixon and Jortner~\cite{BJ} in the late sixties~\cite{PLK}.

Huang \textit{et al} \cite{HSY} have recently discussed an unravelling for the generalized Lindblad equation as applied to the model being discussed in this paper for the case of constant rates. Here, we are interested in the case of time dependent rates involved in the strong coupling regime of this model.

The paper is organized as follows. In Sec.~\ref{sec:s2} we describe the stochastic unravelling of the generalized Lindblad equation that was derived in Ref.~\cite{B1}. In Sec.~\ref{sec:s3} we describe the model used and quote results obtained from the TCL expansion using correlated projection superoperators as derived in Ref.~\cite{BGM}. In Sec.~\ref{sec:s4} we perform the stochastic wave-function simulations for the model and consider both the weak coupling and strong coupling regimes. Results and conclusions follow respectively in the last two sections.

\section{\label{sec:s2}The Generalized Lindblad Equation and its stochastic unravelling}

The general form of the non-Markovian master equation, obtained from the application of correlated projection superoperators, derived in Ref.~\cite{B1} is given by
\begin{equation} \label{gl}
\frac{d}{dt} \rho_i=-\imath [H^i,\rho_i] +\sum_{ j \nu} \left( R^{ij}_\nu \rho_j R^{ij \dagger}_\nu -\frac{1}{2} \left\lbrace 
R^{ji \dagger}_\nu R^{ji}_\nu,\rho_i \right\rbrace \right) 
\end{equation}
where $i,j=1,2, \dots,n$ with $H^i$ being arbitrary Hermitian operators and $R^{ij} $ arbitrary system operators (with $\hbar=1$).
This master equation preserves the normalization and positivity of the density matrix, $\rho_i(t)$.
Following the procedures discussed in Ref.~\cite{BP}, the stochastic unravelling of this equation is obtained by taking $n$ random state vectors $|\psi_i(t)\rangle $ that 
satisfy the stochastic differential equations for a piecewise deterministic process in Hilbert space:
\begin{equation} \label{pdp}
 d|\psi_i\rangle=-\imath G_i |\psi_i\rangle dt + \sum_{j\nu} \left[ \frac{ R^{ij}_\nu |\psi_j\rangle}{\sqrt{M^j_\nu}}-
|\psi_i\rangle\right] d N_\nu^j(t).
\end{equation}
The unnormalized density matrices $\rho_i$ are then determined by the expectation values
\begin{equation}\label{rho}
 \rho_i(t)={\rm E}(|\psi_i(t)\rangle \langle \psi_i(t)|).
\end{equation}

The second term on the right hand side of Eq.~(\ref{pdp}) contains the Poisson increments $dN_\nu^j(t)$ which satisfy,
\begin{equation}
 dN_\nu^j dN_{\nu'}^{j'}=\delta_{\nu \nu'} \delta_{jj'}dN_\nu^j
\end{equation}
and 
\begin{equation}
 {\rm E}(dN_\nu^j)=M^j_\nu dt
\end{equation}
where
\begin{equation}
 M^j_\nu =\sum_i ||R_\nu^{ij}|\psi_j\rangle||^2. 
\end{equation}

The first term on the right hand side of Eq.~(\ref{pdp}) describes the deterministic drift of the process given by
\begin{equation} \label{drift}
 G_i(|\psi_i(t)\rangle )=H^i-\frac{\imath}{2} \sum_{j\nu} \left( R^{ji \dagger}_\nu R^{ji}_\nu - M^j_\nu \right),
\end{equation}
and with this, the deterministic pieces of the process are described by the differential equation
\begin{equation}
 \frac{d}{dt} |\psi_i\rangle = -\imath G_i |\psi_i\rangle.
\end{equation}
The jumps are given by 
\begin{equation}
|\psi_i\rangle \longrightarrow \frac{1}{\sqrt{M^j_\nu}} R^{ij}_\nu |\psi_i\rangle
\end{equation}
which occur at the rate $M^j_\nu$. It should be noted that all state vectors jump simultaneously.

Using the Ito calculus~\cite{BP,CWG} for piecewise deterministic processes, one can demonstrate that the expectation values given by Eq.~(\ref{rho}) satisfies the generalized Lindblad equation (\ref{gl}). The stochastic unravelling nicely illustrates the fact that the master equation preserves the positivity of the $\rho_i$ since an expectation value of the form (\ref{rho}) automatically represents a positive matrix. 

A further remarkable property of the piecewise deterministic process is that the total normalization is strictly preserved under the time evolution:
\begin{equation}
\sum_i\langle \psi_i(t)|\psi_i(t)\rangle \equiv 1.
\end{equation}
This implies that the trace of the reduced density matrix
\begin{equation}
 \rho_S(t)=\sum_i\rho_i(t)
\end{equation}
is strictly conserved (not only on average):
\begin{equation}
 {\rm tr}\rho_S(t)=\sum_i{\rm tr} \rho_i(t)=\sum_i\langle \psi_i(t)|\psi_i(t)\rangle = 1.
\end{equation}
Moreover, the quantities $\langle \psi_i|\psi_i\rangle $ can vary only between $0$ and $1$ and the norm of all components is bounded. This means that there is no exponential growth of the norm of the state vectors as in other Monte Carlo approaches to non-Markovian quantum dynamics.

\begin{figure}[t]
\includegraphics[width=0.4\textwidth]{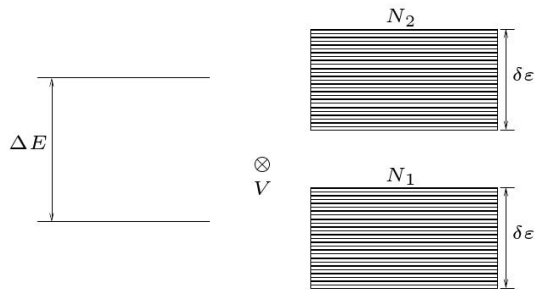}
\caption{{\small A two-state system, with level distance $\Delta E$, coupled to an environment consisting of two energy bands, each  with a finite number of evenly spaced levels $N_1$ and $N_2$. $\delta \varepsilon$ is the width of the bands and $V$ is the system-environment interaction potential.}}
\end{figure}\label{F1}

\section{ \label{sec:s3} The Model and results from the Time-Convolutionless method}

We consider the two-state system coupled to an environment consisting of two energy bands, each  with a finite number of evenly spaced levels. The total Hamiltonian in the Schr\"{o}dinger picture is given by~\cite{BGM},
\begin{eqnarray}\label{Ham}
H &=& \frac{1}{2}\Delta E \sigma_z+\sum_{n-1} \frac{\delta \varepsilon}{N_1}n_1 |n_1\rangle \langle n_1| \nonumber \\ 
&+& \sum_{n_2}\left( \Delta E + \frac{\delta \varepsilon}{N_2} n_2\right)|n_2\rangle \langle n_2| \nonumber \\
&+& V(n_1,n_2) 
\end{eqnarray}
where, the system-environment interaction potential has the form
\begin{eqnarray}
 V(n_1,n_2) = \lambda \sum_{n_1,n_2} c(n_1,n_2)\sp |n_1\rangle \langle n_2| + h.c..
\end{eqnarray}
Here, $n_1/n_2$ labels the levels of the lower($N_1$ levels)/upper($N_2$ levels) energy band and $\lambda$ gives the overall strength of the interaction. $\delta \varepsilon$ is the width of the upper and lower energy bands and $\Delta E$ is the level distance of the two-state system. The coupling constants $c(n_1,n_2)$ are complex Gaussian random variables with zero mean and unit variance.\\

We consider the initial state where only the lower band is occupied. For the weak coupling case where $\delta \epsilon t \gg 1$, the second order of the TCL expansion using correlated projection superoperators, which we call new TCL2, gives the following equations of motion~\cite{BGM}:
\begin{eqnarray} 
\frac{d}{dt} \rho_1 &=& \go \sp \rho_2 \sm - \frac{\gt}{2} \{ \sp \sm,\rho_1\} \label{mod1}\\
\frac{d}{dt} \rho_2 &=& \gt \sm \rho_1 \sp - \frac{\go}{2} \{ \sm \sp,\rho_2\} \label{mod2},
\end{eqnarray}
with the following solution for the population of the upper level,
\begin{equation}\label{newtcl}
 \rho_{11}=\rho_{11}(0)\left[ \frac{\go}{\go+\gt}+ \frac{\go}{\go+\gt} \text{e}^{-(\go+\gt)t}  \right].
\end{equation}

For the case where the times $t$ do not satisfy the condition $\delta \epsilon t \gg 1$ (strong coupling), the second order of the TCL expansion using correlated projection superoperators, which we call new TCL2(t), the equations of motion are:
\begin{eqnarray}
\frac{d}{dt} \rho_1 = \int_0^t dt_1 h(t-t_1) [ 2\go \sp \rho_2 \sm \nonumber \\- \gt \{ \sp \sm,\rho_1\} ]  \label{modt1}, \\
\frac{d}{dt} \rho_2 = \int_0^t dt_1 h(t-t_1) [ 2\gt \sm \rho_1 \sp \nonumber \\ - \go \{ \sm \sp,\rho_2\} ] \label{modt2},
\end{eqnarray}
where $\gt h(t-t_1)$ is the two-point environment correlation function such that
\begin{equation}
 h(t)=\frac{\delta \varepsilon}{2 \pi} \frac{\sin^2(\delta \varepsilon t/2)}{(\delta \varepsilon t/2)^2}.
\end{equation}
The solution for the populations of the upper level in this case is given by,
\begin{equation} \label{newtcl_t}
 \rho_{11}=\rho_{11}(0)\left[ \frac{\go}{\go+\gt}+ \frac{\go}{\go+\gt} \text{e}^{-\Gamma(t)}  \right],
\end{equation}
where
\begin{equation}
\Gamma(t)=2(\go+\gt)\int_o^t dt_1 \int_0^{t_1} dt_2 h(t_1-t_2).
\end{equation}

For both cases, the relaxation rates are given by,
\begin{equation}
 \gamma_{1,2}=\frac{2 \pi \lambda^2 N_{1,2}}{\delta \varepsilon}.
\end{equation}

\section{\label{sec:s4} Stochastic Wave-Function Simulations}
In this section we perform Monte Carlo simulations of the generalized master equation for our model for both the weak coupling and strong coupling cases. The terminology, weak coupling and strong coupling are used in the same sense as described in Ref.~\cite{BP}. Details of the simulation algorithm can also be found in Ref.~\cite{BP}. 

\subsection{Weak Coupling}
It is clear to see that Eqs.~(\ref{mod1}) and~({\ref{mod2}) are of the same form as Eq.~(\ref{gl}) with the associations
\begin{eqnarray}
H^1=H^2=0,  &&  R^{11}=R^{22}=0, \\
R^{12}=\sqrt{\go} \sp,  &&  R^{21}=\sqrt{\gt} \sm.
\end{eqnarray}
Here we have $n=2$ and therefore consider two state vectors $|\psi_1\rangle$ and $|\psi_2\rangle$. The drift terms for the model from 
Eq. (\ref{drift}) are therefore given by
\begin{eqnarray}
G_1 &=& -\frac{\imath}{2}(\gt\sp\sm -\gt||\sm | \po \rangle ||^2\cdot \mathbbm{1} - \go||\sp | \pt \rangle ||^2\cdot \mathbbm{1}) \nonumber \\
&=& -\frac{\imath}{2}
\begin{pmatrix}
\gt - \gt c_1 -\go c_2 & 0  \\
0 & -\gt c_1 - \go c_2
\end{pmatrix},
\end{eqnarray}
with realizations
\begin{equation}
 |\po(t) \rangle \to \frac{\text{e}^{-\imath G_1 t}|\po \rangle}{||\text{e}^{-\imath G_1 t}|\po \rangle||}
\end{equation}
and
\begin{eqnarray}
G_2 &=& -\frac{\imath}{2}(\go\sm\sp -\gt||\sm | \po \rangle ||^2\cdot \mathbbm{1} - \go||\sp | \pt \rangle ||^2\cdot \mathbbm{1}) \nonumber\\
&=& -\frac{\imath}{2}
\begin{pmatrix}
- \gt c_1-\go c_2 & 0  \\
0 & \go -\gt c_1 - \go c_2
\end{pmatrix}
\end{eqnarray}
with realizations
\begin{equation}
 |\pt(t) \rangle \to \frac{\text{e}^{-\imath G_2 t}|\pt \rangle}{||\text{e}^{-\imath G_2 t}|\pt \rangle||},
\end{equation}
where $c_1=||\sm | \po \rangle ||^2$ and $c_2=||\sp | \pt \rangle ||^2$.

The two possible jumps are 
\begin{equation}
|\po\rangle \to 0,~~|\pt\rangle \to \frac{\sm |\po\rangle}{||\sm |\po\rangle||}
\end{equation}
with rate $M^1=\gt  ||\sm |\po\rangle||^2 $ and 
\begin{equation}
 |\po\rangle \to \frac{\sp |\pt\rangle}{||\sp |\pt\rangle||},~~|\pt\rangle \to 0
\end{equation}
with rate $M^2=\go  ||\sp |\pt\rangle||^2 $.

\begin{figure}[t] 
\includegraphics[width=0.48\textwidth]{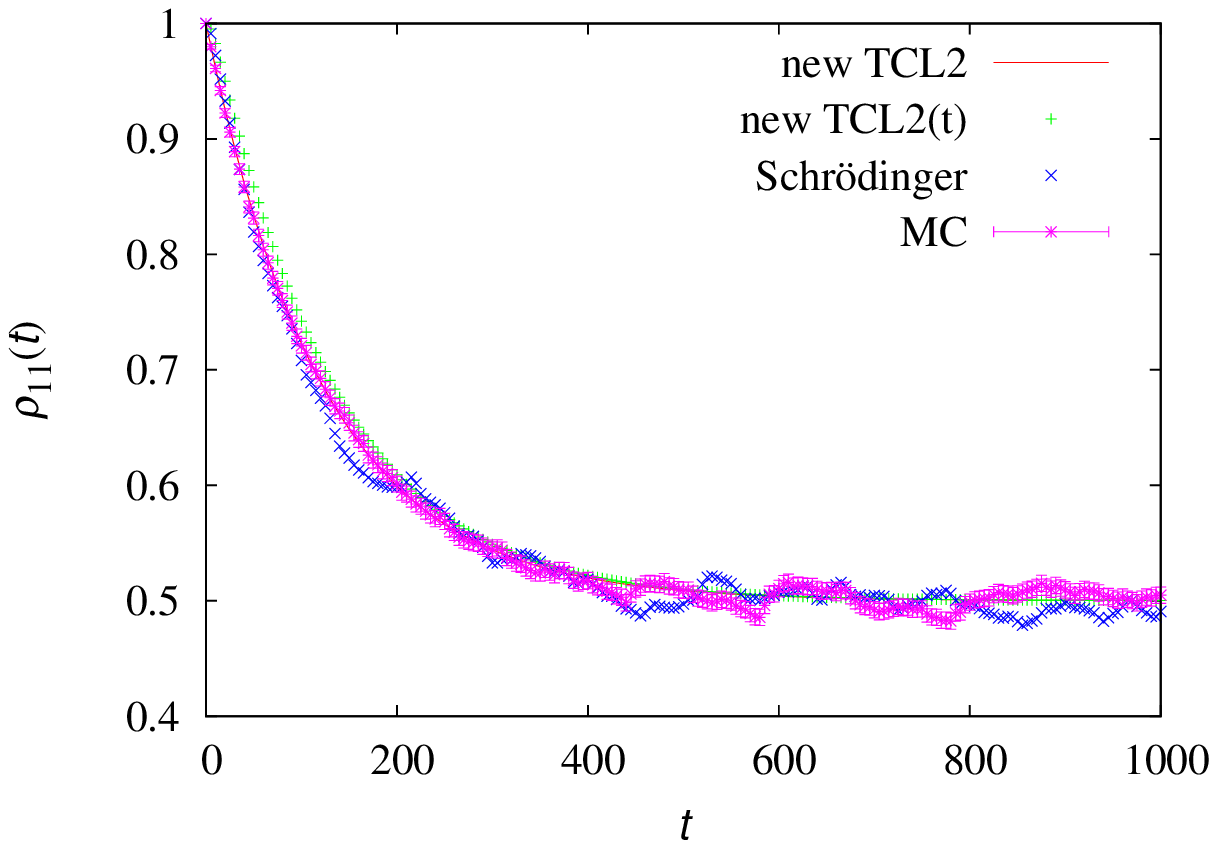}
\caption{\label{fig1} (Color online) Comparison of the four methods with $N_1=N_2=200,~\delta \epsilon=0.31$ and $\lambda=0.001$. 'new TCL2' and 'new TCL2(t)' correspond to Eq. (\ref{tcl2}) and Eq. (\ref{tcl2t}) respectively. The Monte Carlo simulation, 'MC', was done with time independent rates and the 'Schr\"{o}dinger' gives the exact result.}
\end{figure} 

The total waiting time distribution is
\begin{eqnarray}
F(\tau)&=&1- \exp[- \sum_{ij,\lambda} ||R^{ij}_\lambda |\psi_j \rangle||^2 \tau ] \nonumber\\
&=& 1-\exp[-\go||\sp|\pt \rangle ||^2 \tau - \gt||\sm|\po \rangle ||^2 \tau] \nonumber\\
&=& 1- \exp[-\go c_2 \tau- \gt c_1 \tau]
\end{eqnarray}
and depending on the current realizations, $c_1$ or $c_2$ equals zero. It is easy to see that this process is rather simple, in that, beginning with the initial state $|\po(0) \rangle=|e \rangle$ and $|\pt(0)=0 $, the process simply jumps between $|\po \rangle=|e \rangle,~|\pt\rangle=0 $ and $|\po\rangle=0,~ |\pt\rangle=|g \rangle $.

\subsection{Strong Coupling}

In this case, Eqs.~(\ref{modt1}) and~({\ref{modt2}) are of the same form as Eq.~(\ref{gl}) with the associations
\begin{eqnarray}
H^1=H^2=0,  &&  R^{11}=R^{22}=0, \\
R^{12}=\sqrt{2\go} \sp,  &&  R^{21}=\sqrt{2\gt} \sm.
\end{eqnarray}
The drift terms and realizations are of the same form as for the weak coupling case, except here we need to take into consideration the time dependence in the waiting times. The total waiting time distribution is given by
\begin{widetext} 
\begin{eqnarray}
 F(\tau)&=&1- \exp[2\int_0^\tau dt_1 h(\tau-t_1) (-\go c_2 \tau- \gt c_1 \tau)] \nonumber \\
&=& 1- \exp\left[  \frac{2(-1 +\cos(\delta \varepsilon \tau) + \delta \varepsilon \tau~ \text{Si}(\delta \varepsilon \tau))}{\delta \varepsilon \tau \pi}(-\go c_2 \tau- \gt c_1 \tau)\right],
\end{eqnarray}
\end{widetext}
where $\text{Si}(\omega)=\int^\omega_0 \frac{\sin x}{x} dx$. Once again, depending on the current realizations, $c_1$ or $c_2$ equals zero.

\begin{figure}[t] 
\includegraphics[width=0.48\textwidth]{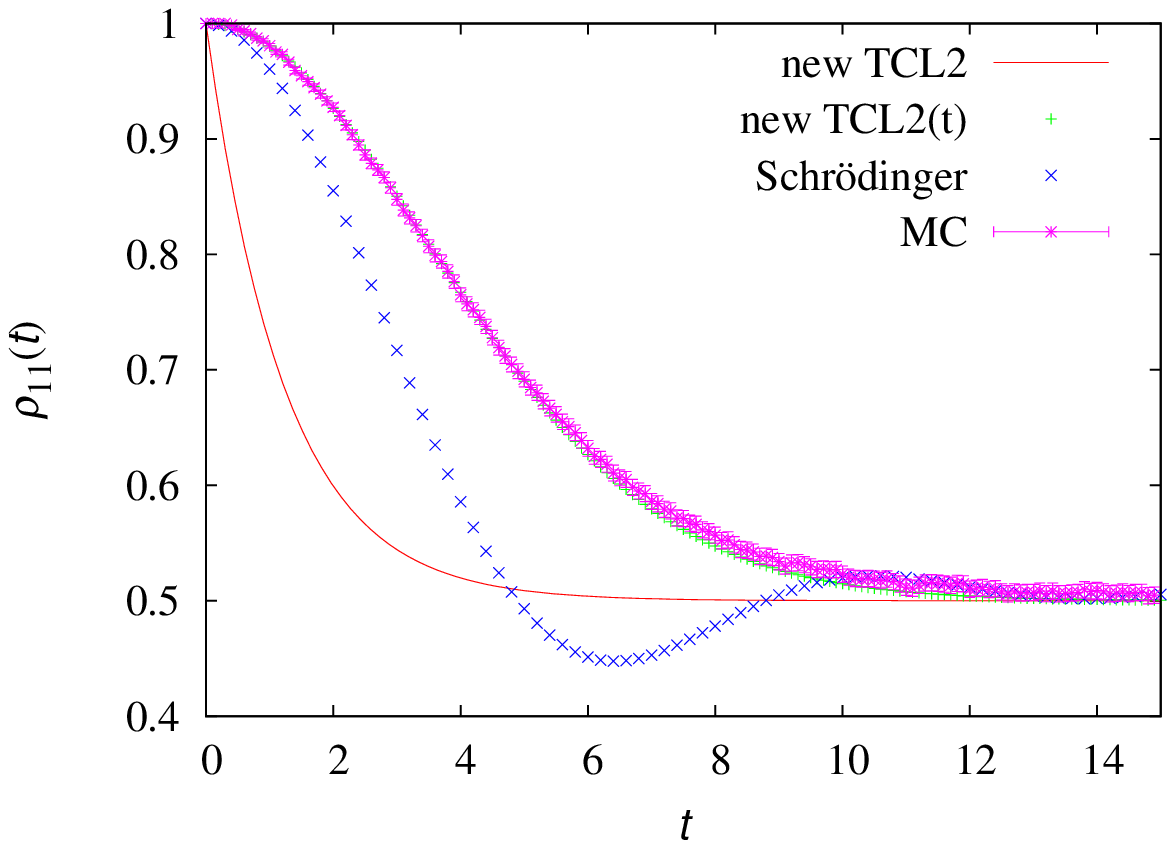} 
\caption{\label{fig2} (Color online) Comparison of the four methods with $N_1=N_2=200,~\delta \epsilon=0.31$ and $\lambda=0.01$. 'new TCL2' and 'new TCL2(t)' correspond to Eq. (\ref{tcl2}) and Eq. (\ref{tcl2t}) respectively. The Monte Carlo simulation, 'MC', was done with time dependent rates and the 'Schr\"{o}dinger' gives the exact result.}
\end{figure} 

\section{ Results}

In both cases we have considered the environment with $N_1=N_2=200$ energy levels and the relaxation rates $\gamma=\go=\gt$. $\delta \varepsilon $ was chosen to be $0.31$ so that for $\lambda=0.001$, the ratio $\frac{\gamma_{1,2}}{\delta \varepsilon}=0.013$ and for $\lambda=0.01$, $\frac{\gamma_{1,2}}{\delta \varepsilon}=1.3$. Note that for the two cases considered, the relaxation rates differ by a factor 100.

For the simulation of the new TCL2 with time-independent rates the waiting time distribution is $F(\tau_{1,2})=\exp(-\gamma_{2,1} \tau_{1,2})$, which is just the exponential distribution. For the initial condition $\rho_{11}(0)=1$, we simulate 
\begin{equation} \label{tcl2}
 \rho_{11}(t)=\frac{1}{2}+\frac{1}{2}\rm{e}^{-2\gamma_{1,2} t}.
\end{equation}
For the simulation of the new TCL2(t), the procedure is the same except that we need to include the time dependence in the waiting times. A Gaussian quadrature algorithm was used to evaluate the integral of $h(\tau-t_1)$ and a polynomial interpolation algorithm was used to extract the waiting times, $\tau_{1,2}$. With initial condition $\rho_{11}(0)=1$, we simulate
\begin{equation}\label{tcl2t}
\rho_{11}(t)=\frac{1}{2}+\frac{1}{2}\text{e}^{-\Gamma_{1,2}(t)}, 
\end{equation}
\\
where $\Gamma_{1,2}(t)=4 \gamma_{1,2} \int^t_0 dt_1 \int^{t_1}_0 dt_2 h(t_1-t_2)$. In both cases the Monte Carlo simulations were done with the initial state: $|\po(0) \rangle=|e \rangle$ and $|\pt(0) \rangle=0$. Also, in both cases, 5000 trajectories where used in the Monte Carlo simulations to recover the quantum master equation.

We have also performed numerical solutions of the full Schr\"{o}dinger equation corresponding to the Hamiltonian given in Eq. (\ref{Ham}). The initial state was taken to be $|1 \rangle \otimes |\chi \rangle$, where the environmental state $|\chi \rangle$ was of the form
\begin{equation}
\langle \chi | = (\overbrace{0, \dots \dots, 0}^{N_2}, d_1, \dots \dots, d_{N_1}),
\end{equation}
where $d_1, \dots, d_{N_1}$ are Gaussian random variables with zero mean and variance equal to one. $\Delta E$, the level distance of the two-state system was taken to be unity.

In Figs.~{\ref{fig1}} and {\ref{fig2}} we compare the results of the four different methods discussed in the paper, i.e., the new TCL2 given by Eq.~(\ref{tcl2}), the new TCL(t) given by Eq.~(\ref{tcl2t}), the numerical solution of the Schr\"{o}dinger equation and the Monte Carlo simulations based on the unravelling of the master equation.
For the weak coupling, Fig.~{\ref{fig1}} shows a good overlap of all four methods. For the strong coupling, as shown in Fig.~{\ref{fig2}}, the Monte Carlo simulation results overlap almost completely with the new TCL2(t) method and also gives the correct stationary state and relaxation time when compared to the exact result obtained by solving the Schr\"{o}dinger equation.

\section{ Conclusions}
In this paper, we have performed a stochastic unravelling of the generalized Lindblad master equation~\cite{B1} and applied it to a two-level system coupled to an environment consisting of two energy bands with 200 energy levels each. Our unravelling was applicable to both the weak coupling regime with time independent rates and the strong coupling regime with time dependent rates, for this model. Our Monte Carlo simulation results were found to be in good agreement with the second order time-convolutionless projection operator method results as obtained by the authors of Ref.~\cite{BGM}.

\section*{ACKNOWLEDGMENTS}
The authors would like to acknowledge insightful discussions with Heinz-Peter Breuer without whose stimulating input this work would not have been done. This work is based upon research supported by the South African Research Chair Initiative of the Department of Science and Technology and National Research Foundation. M.M and F.P also acknowledge financial support and the use of the facilities at the Centre for High Performance Computing (CHPC).

\end{document}